\documentclass[superscriptaddress,showpacs,aps,twocolumn,floatfix,prb,longbibliography]{revtex4-2}
\usepackage{bm,amsmath,amssymb}
\usepackage{amssymb}
\usepackage{amsmath}
\usepackage{commath}
\usepackage{graphicx,bm}
\usepackage{verbatim}
\usepackage[latin1]{inputenc}
\usepackage[dvipsnames]{xcolor}
\usepackage{soul}
\usepackage[separate-uncertainty = true]{siunitx}
\usepackage{ bbold }
\usepackage[colorlinks]{hyperref}
\AtBeginDocument{%
    \hypersetup{    
        linkcolor=blue,
        filecolor=magenta,      
        citecolor=blue,
        colorlinks=true,
     }
}

\begin{document}

\title{Signatures of Jackiw-Rebbi resonance in the thermal conductance of topological Josephson junctions with magnetic islands}

\author{Daniel Gresta}
\thanks{These two authors contributed equally. dgresta@unsam.edu.ar }
\affiliation{International Center for Advanced Studies, ECyT-UNSAM,  25 de Mayo y Francia, 1650 Buenos Aires, Argentina}
\author{Gianmichele Blasi}
\thanks{These two authors contributed equally. gianmichele.blasi@sns.it }
\affiliation{NEST, Scuola Normale Superiore and Instituto Nanoscienze-CNR, I-56126, Pisa, Italy}
\author{Fabio Taddei}
\affiliation{NEST, Scuola Normale Superiore and Instituto Nanoscienze-CNR, I-56126, Pisa, Italy}
\author{Matteo Carrega}
\affiliation{SPIN-CNR, Via Dodecaneso 33, 16146 Genova, Italy}
\author{Alessandro Braggio}
\affiliation{NEST, Scuola Normale Superiore and Instituto Nanoscienze-CNR, I-56126, Pisa, Italy}
\author{Liliana Arrachea}
\affiliation{International Center for Advanced Studies, ECyT-UNSAM,  25 de Mayo y Francia, 1650 Buenos Aires, Argentina}

\begin{abstract}
Josephson junctions in two-dimensional topological insulators with embedded magnetic domains can host a number of topological phases, in particular, Jackiw-Rebbi solitons and Majorana zero modes. These different 
non-trivial phases appear in such junctions  for multiple-domain magnetic islands, showing a rich multi-gap structure. Features related to the interplay between superconductivity and magnetism in these systems cannot be easily discerned looking at
behavior of the Andreev spectrum and the concomitant dc Josephson effect. Instead, the thermal conductance is very sensitive to the nature
of the junction and the domain structure of the magnetic island. 
We present a detailed analysis of these properties in the case of a topological Josephson junction with a single and two-domain magnetic island.
Configurations hosting soliton magnetic modes lead to a peculiar behavior of the thermal conductance relative to the thermal quantum, characterized by a negative slope as a function of the temperature, just above the superconducting critical temperature. At low temperatures, these junctions also show characteristic coherence patters in the behavior of the thermal conductance as function of the Josephson phase bias and the angle between the magnetizations of the domains.  
\end{abstract}

\maketitle

\section{Introduction.}
Heterostructures based on two-dimensional topological insulators (2DTI) 
have received a great attention in the last years for the interesting physics and promising applications
\cite{heli1,heli2,heli3,heli4}. 
The edge states defining helical one-dimensional conducting channels
offer a large variety of quantum phenomena in combination with nanomagnets\cite{barash2002,duan2015,ghosh2017} and superconductors\cite{stanescu2010proximity,sedlmayr2020hybridization,zhang2019,Michelsen2020,ronetti2017,ronetti2020}.

A prominent example is the platform for topological superconductivity proposed by Fu and Kane\cite{fu-kane}, which consists in a Josephson junction made of a Kramers pair of
helical edge states in close proximity to a s-wave superconductor and a magnet embedded in the junction. 
For a magnetic moment having a component perpendicular to the natural
quantization axis of the 2DTI, Majorana bound states are formed. The concomitant signatures in the behavior of the Josephson current have been investigated in several works\cite{fu-kane,meng,jiang,houzet,barbarino2013,kacho,yacoby,crepin,keideltun,hart,laurens,blasi2019,ren2019,marra2016}.

The key role played by a magnetic island placed inside the 2DTI is to introduce a boundary with a backscattering process in the Dirac system constituted by the helical edge states. Without the superconducting ingredient, this phenomenon leads to interesting effects in the electron  transport \cite{qi,meng1,nanom,piet,patrick1,patrick2} and in thermoelectric\cite{Roura-Bas2018Nov,optimal,hajiloo} properties. The fact that the magnetic island may have multiple domains further extends  the scenario to interesting topological structures. The simplest of such situations corresponds to two domains with opposite orientations of the magnetic moments, which 
 is a realization of the so-called Jackiw-Rebbi (JR) model of a one-dimensional Dirac system with a space-dependent soliton mass \cite{jackreb,shen2012}. Similarly to the discrete Hamiltonian by Su-Schrieffer-Heeger\cite{ssh}(SSH), this model is known to host topologically protected modes within the spectral gap. The high thermoelectric response generated as a consequence of these modes was recently pointed out in Ref. \onlinecite{optimal}. 
 JR physics on junctions with embedded superconductors was recently addressed also in Refs. \onlinecite{ziani,mora}, 
 while the emergence of other states with fractional charges in helical edge states  with many-body interactions has been also studied\cite{maciejko,zhang-kane,ziani-2015}.
 
 \begin{figure}[htp]
	\centering
	\includegraphics[width=\columnwidth]{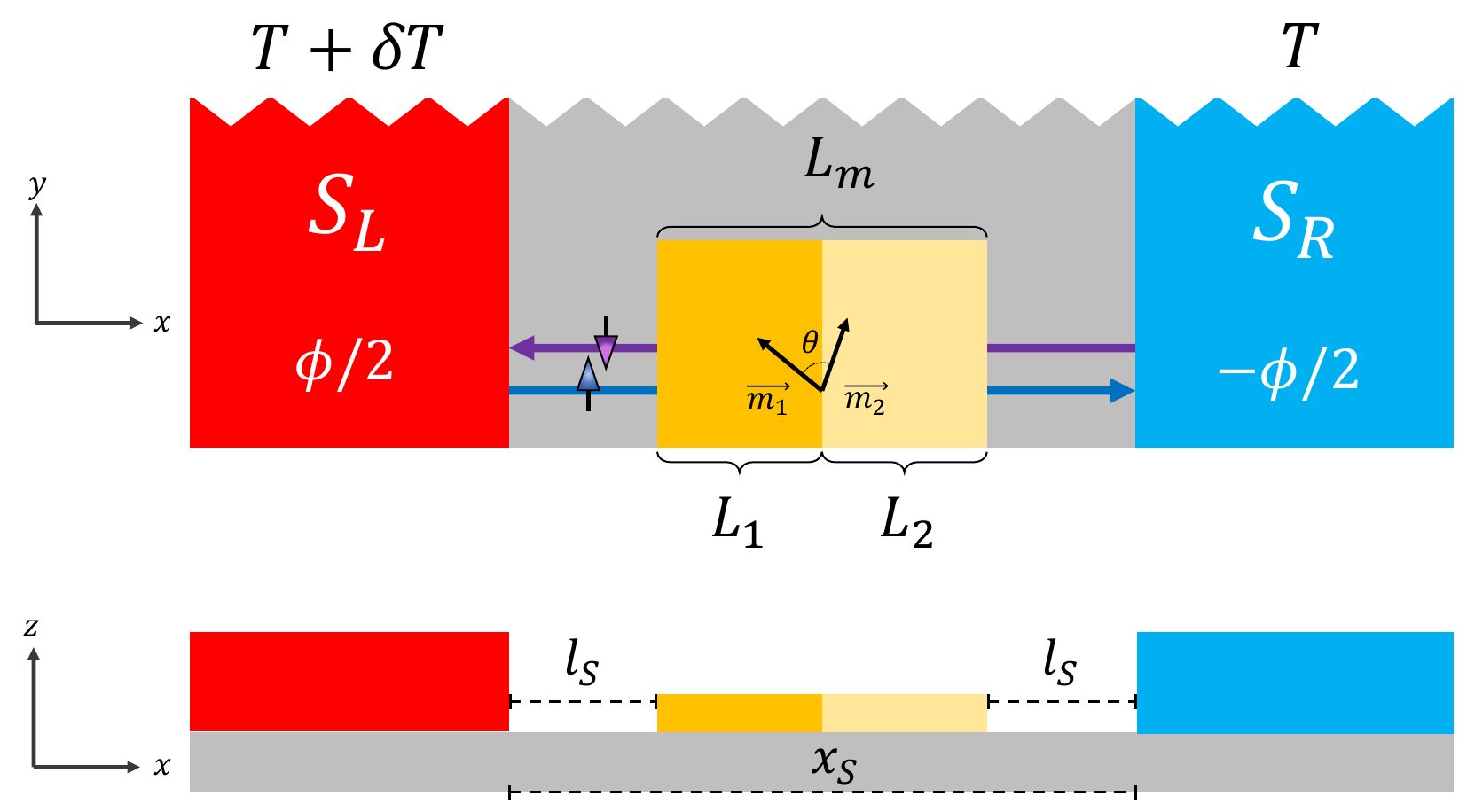}
	\caption{Top and lateral view of the device. Two
	semi infinite
	superconductors at slightly different temperatures, $T$ and $T+\delta T$ and with a phase bias $\phi$, proximized to a 2DTI in the quantum spin Hall regime.
	A magnetic island, of total length $L_m$ composed with two magnetic domains of length $L_1$ and $L_2$ respectively, is placed at distance $l_S$ from both superconductors, and contacted to the Kramers pair of helical states localized at one of the edges.   
	The magnetic moments of the two domains are oriented with a relative tilt $\theta$.}
	\label{fig:sketch}
\end{figure}

Thermal and thermoelectric effects in 2DTI in contact with superconductors have recently attracted a significant interest\cite{soth-han1,soth-han2,shapi,keidel,nonloc1,nonloc2,bours2018,hwang2020,zhang2017manipulation,shaf2020}. 
It is remarkable that topological properties, which typically have associated spectral features close to zero energy may also have an impact on the
thermal response. In this sense, interference patterns in topological Josephson junctions were studied in Refs.~\onlinecite{soth-han1,nonloc2}. The aim of the present work is to analyze the thermal conductance of a topological Josephson junction with an embedded magnetic island. 
A sketch of the  device  is shown in Fig. \ref{fig:sketch}, which consists of a Josephson junction constructed by proximity effect to a 2DTI with a magnetic island contacting the two states of the Kramers pair in one of the edges. The junction is biased with a small temperature difference, $\delta T$ and with  a phase difference, $\phi$, between the two superconducting pairing potentials. 
Here, we will focus on a magnetic island with one or two magnetic domains.
Our goal is to identify features in the thermal transport that could indicate the topological nature of the junction.
We show that the thermal conductance is very sensitive to the characteristics of the junction, in particular, to the domain structure of the magnetic island.
Interestingly, systems hosting JR resonant states lead to a peculiar behavior of the thermal conductance, such that it decreases for increasing temperature just above the 
superconducting critical temperature.

The work is organized as follows. In Section \ref{sec:model} we present the model for the Kramers pair of 2DTI edge states in contact to s-wave superconductors with a phase bias and magnetic domains with different orientations of the magnetic moments.  In Section \ref{sec:scat} we discuss the scattering matrix approach used to analyze the topological junction. In Section \ref{sec:results} we present our main results concerning the Andreev spectrum and the thermal conductance obtained in different configurations. Section \ref{sec:con} is devoted to summary and conclusions.

\section{Model}\label{sec:model}
The system under investigation is depicted in Fig. \ref{fig:sketch}. It consists of a 2DTI strip attached to two  superconducting electrodes with a phase difference $\phi$ are
 placed on top and  kept at slightly different temperatures, $T, ~T+\delta T$ (see light blue and red blocks in the sketch). Due to the proximity effect, the two superconductors induce a pairing potential in the portion of the 2DTI beneath it. In addition, a magnetic island with one or two domains (yellow blocks) with the magnetization directions forming 
an angle $\theta$, are put in contact with one of the pairs of edge states of the strip. The lengths of the two magnetic domains along the edge are $L_1$ and $L_2$, respectively, and they are placed at distance $l_S$ from each superconducting electrode. 
The width of the TI strip is assumed to be large enough such that the helical states
(represented by solid lines) on the two edges are uncoupled one another and therefore we can restrict our analysis to a single Kramers pair.
The Hamiltonian 
describing the system taking into account
the proximity-induced pairing potential and the coupling to the magnetic island,
expressed in the basis of Nambu spinors $\Psi(x)= \left(\psi_{\uparrow}(x), \psi_{\downarrow}(x),\psi^{\dagger}_{\downarrow}(x), -\psi^{\dagger}_{\uparrow}(x) \right)^T$,
reads 
\begin{equation}\label{ham}
H= \int_{-\infty}^{+\infty} dx \Psi^{\dagger}(x) \left[ {\cal H}_0(x)+ {\cal H}_M(x) + {\cal H}_S(x) \right]   \Psi(x).
\end{equation}
The term
\begin{equation}\label{h0}
{\cal H}_0(x)= \left(- i \hbar v_F \partial_x \right) \sigma_z \tau_z -\mu\sigma_0 \tau_z,
\end{equation}
describes the free Kramers pair, with $z$ as the natural quantization axis of the topological insulator. The terms 
\begin{equation}\label{hmdel}
{\cal H}_M(x)= J \vec{m}(x) \cdot \vec{\sigma} ,\;\;\;\;\;\;\;\;\;\;\;\;\;\; {\cal H }_S= \vec{\Delta}(x) \cdot \vec{\tau},
\end{equation}
describe, respectively, the effect of the coupling to the magnetic island and the BCS superconducting potential. The matrices $\sigma_0, \vec{\sigma}=(\sigma_x,\sigma_y,\sigma_z)$ and
$\tau_0, \vec{\tau}=(\tau_x,\tau_y,\tau_z)$ operate, on the spin and particle-hole degrees of freedom, respectively.
The pairing potential induced by superconducting proximity is described as follows,
\begin{equation}\label{delta}
    \vec{\Delta}(x) = 
    \left[\vec{\Delta}(\phi/2) \Theta(-x) +
\vec{\Delta}(-\phi/2)\Theta(x-x_{S}) \right] ,
\end{equation}
with $\vec{\Delta}(\pm \phi/2)= \Delta_0 \left(\cos \phi/2, \pm \sin \phi/2,0 \right)$ where $\phi$ is the phase bias and $x_S$ is the distance between the two superconductors,
which are considered to be semi-infinite
The magnetization of the island is accounted for by
\begin{eqnarray}\label{magnet}
\vec{m}(x) &=& \vec{m}_1  \left[ \Theta(x-l_S) - \Theta(x-x_1)  \right] \nonumber \\
& + & \vec{m}_2  \left[ \Theta(x-x_1) - \Theta(x-x_2)  \right],
\end{eqnarray}
where $\vec{m}_j=m_j \left( \cos \theta_j, \sin \theta_j, 0 \right)$, for $j=1,2$ and $x_1=L_1 +l_S $ and $x_2=L_2+x_1$. The total length of the magnetic island is
$L_{m}=L_1+L_2$.
The case with $\theta_1=\theta_2$ and $m_1=m_2 \equiv m$ effectively reduces to a single magnetic domain of length $L_m$. 
For sake of simplicity, we assume 
a fully anisotropic magnetic moment with a  vanishing $\hat z$-component of the magnetization (direction parallel to the natural quantization axis of the topological insulator). 
Notice that the component of the magnetization perpendicular to $\hat z$  is the only mechanism introducing a backscattering processes in the present problem, and it is precisely this ingredient the one generating non-trivial effects in 
 the two-terminal transport properties. 
For simplicity we discuss
results corresponding to the configuration where the magnetic island occupies all the space of the junction, in which case $l_S=0$ and $x_S=L_m$. 
However, for $l_S \neq 0$, we find qualitatively similar features.
 
In the absence of superconducting contacts ($\Delta=0$), the component of the magnetization perpendicular  to the natural spin quantization axis of the helical edge states is analogous to
a mass term in the Dirac system and opens a gap in the spectrum. There are several consequences when this mass is not uniform in space. 
The magnetic island indeed plays the role of a barrier for the propagating helical states and 
it must be long enough in order to completely suppress the tunneling\cite{optimal}.
Hence, in order to have a well defined gap in the spectrum, for which the tunneling probability is exponentially suppressed, the magnetic island must be larger than the characteristic length
\begin{equation}
\xi_M =  \frac{\hbar v_F}{Jm},
\label{eq:xim}
\end{equation}
being $Jm$ the magnetic energy gap in the limit of uniform magnetization along infinite-length helical modes.
In summary, the inequality $L_m > \xi_{M}$
must be satisfied in order to show a clear suppression of the transmission probability due to the opening of the magnetic gap. 
Another interesting effect introduced by a non-uniform magnetization takes place in the case of
two magnetic domains with exactly opposite orientations, i.e. $\theta=\pi$, which realizes 
the JR model\cite{jackreb}, where two consecutive masses with different sign define a soliton domain wall  in a one-dimensional Dirac system. 
This model, along with its SSH discrete version\cite{ssh} hosts a topological zero mode at the interface. As analyzed in Ref. \onlinecite{optimal},
this mode survives as a resonant state
in the magnetic gap, shifted away from zero energy for a wide range of relative tilting angle $\theta$ in the orientations of the magnetic moments, and the width of this resonance scales with the inverse of the length of the magnetic domains. 
The reason for its stability is due to the fact that the system realizes a Goldstone-Wilczek soliton, as discussed in Refs. \onlinecite{goldstone,hughes,flecks}.

In combination with superconductivity, for a finite magnetization embedded in the junction between the two superconductors, a
topological state develops, with Majorana zero modes localized in the boundaries between the superconductors and the magnetic island \cite{fu-kane,jiang,crepin}. 
Interestingly, a magneto-Josephson duality exists \cite{jiang}, such that the role of the magnetization can be interchanged with the superconducting potential. This can be understood by noticing that in the Hamiltonian for the device, Eq.(\ref{hmdel}), the terms with the Pauli matrices acting on the spin degrees of freedom, $\vec \sigma$, have the same
structure as those with Pauli matrices $\vec \tau$, which act on the particle-hole degrees of freedom. Due to the s-wave nature of the superconducting order parameter, the
relevant physical parameter characterizing the orientation of the magnetic moments is the relative tilt $\theta$. Furthermore, this angle is related through the above mentioned duality to the
phase difference $\phi$ between the two superconductors.   

Importantly, in the presence of the superconducting contacts, the other characteristic length in the problem is the superconducting  coherence
length $\xi_S = \hbar v_F/\Delta$. As we will see, most of the interesting effects in the behavior of the thermal conductance arise from the interplay between the
magnetic and superconducting spectral gaps. The conditions under which they are most remarkable correspond to comparable values for the two characteristic lengths $\xi_M$ and $\xi_S$.

\section{Scattering matrix approach}\label{sec:scat}
We rely on the scattering matrix approach to evaluate the subgap Andreev spectrum, as well as the transmission function ruling the behavior of the thermal conductance. 
In the absence of inelastic processes, dc transport is determined by the quantum mechanical matrix $S$, which yields scattering properties at energy $\varepsilon$, of a phase-coherent, non-interacting system described by the Hamiltonian $H$ of Eq. (\ref{ham}). 
The scattering problem in terms of the $S$-matrix can be formulated as
\begin{equation}\label{eq: scattering matrix}
   \left. \Psi^{\alpha}_{(i,\sigma)}\right |_{\mathrm{out}} = S^{\alpha,\beta}_{(i,\sigma)(j,\sigma')} \left. \Psi^\beta_{(j,\sigma')} \right|_{\mathrm{in}} ,
\end{equation}
where summation is implicit on repeated indices.
This equation relates incoming/outgoing states 
$(j,\sigma')/(i,\sigma)$ with $\left\{ \sigma,\sigma' \right\}=$ 
$\left\{ \uparrow,\downarrow\right\}$ 
labeling the spin-channel at the respective superconducting lead $i,j = L,R$.
In Eq.(\ref{eq: scattering matrix}), $\left\{\alpha,\beta\right\}=\left\{\tilde {e},\tilde{h}\right\}$ label the quasiparticles (QPs) and quasiholes (QHs) in the superconductors. Following the standard procedure presented in Ref. \onlinecite{Datta-book}, we computed the full scattering matrix of the system
\begin{equation}\label{eq: sm combinated main}
    S = S_L\circ S_M \circ S_R.
\end{equation}
The matrices $S_{L,R}$ describe the left and right interfaces of the 2DTI with the superconductors. These matrices are combined  with the matrix $S_M$ describing the 2DTI edges in contact with the magnetic domain. In Appendix \ref{sec: Appendix} we present in more detail the calculation of the different  matrices  $S_{L,R}$ (Sec. \ref{sms}) and $S_M$ (Sec. \ref{smm}). By taking the trace over spin channels of the scattering matrix of Eq. (\ref{eq: sm combinated main}) we can compute the probability scattering coefficients\cite{lambert1998}
\begin{equation}\label{eq: Pmatrix}
    P_{i,j}^{\alpha,\beta} = \sum_{\sigma,\sigma'}\left|S^{\alpha,\beta}_{(i,\sigma),(j,\sigma')} \right|^2,
\end{equation}
which represents the reflection $(i=j)$ or transmission $(i\neq j)$
probabilities of a quasiparticle of type $\beta$ in the lead j to a quasiparticle of type $\alpha$ in lead $i$.

\subsection{Andreev bound states and thermal conductance}
Under suitable conditions \cite{beenakker1,martin2011josephson},
Andreev bound states develop with energies below the superconducting gap $\Delta$. These states are crucial in the behavior of the Josephson current.  
To calculate the Andreev spectrum for $|\varepsilon|<\Delta$, we proceed as in Refs. \onlinecite{crepin,beenakker1}. 
Recall that for the non-superconducting region we have 
$\Psi_{\mathrm{out}} = S_M \Psi_{\mathrm{in}}$, with $\Psi_{\mathrm{in,out}}$  defined as in Eq. (\ref{eq: scattering matrix}) (see also Eq. (\ref{eq: SM})). For the subgap regime only perfect Andreev reflection is allowed, which leads to $\Psi_{\mathrm{in}} = S_A\Psi_{\mathrm{out}}$, being
\begin{align}
 S_A  =& \exp\left[2i\frac{\varepsilon}{\Delta}\frac{l_S}{\xi_S} -i\mathrm{\arccos} (\frac{\varepsilon}{\Delta}) \right ]\times \nonumber \\ 
 & \begin{pmatrix}
0 & \mathrm{Diag}[e^{i\phi/2},e^{-i\phi/2}] \\
\mathrm{Diag}[e^{-i\phi/2},e^{i\phi/2}] & 0
\end{pmatrix}.
\end{align}
By combining these two expressions we obtain the so called compatibility equation  \cite{beenakker1} 
\begin{equation}
    \mathrm{Det}\left[1-S_A S_M \right] = 0,
\label{eq: compatibility}
\end{equation}
which we solve numerically to investigate the Andreev bound state spectrum.

Given the scattering matrix, we can also calculate the heat current generated as a response to the temperature bias $\delta T$.
This quantity provides a complementary information to the Andreev spectrum, since it depends only on the quasiparticle spectrum above the gap. We focus on small $\delta T$, such that linear response applies. 
Indeed the corresponding heat current in this regime is proportional to $\delta T$ 
and the thermal conductance is the natural transport coefficient.
We find it convenient to characterize the response to the thermal bias in terms of 
the {\em relative thermal conductance} $\kappa_{\rm th} (T)$, 
which is expressed
with respect to the quantum of thermal conductance $G_T = \pi^2 k_B^2T/3h$. It reads \cite{butcher1990thermal,benenti2017fundamental,lambert1998}
\begin{equation}\label{eq:kappa}
    \kappa_{\rm th} (T) = -\frac{1}{G_T}\int_\Delta^\infty \varepsilon^2 \frac{\partial f (\varepsilon) }{\partial\varepsilon} {\cal T} (\varepsilon)  d\varepsilon;
\end{equation}
where $f(\varepsilon) = \left [\exp\left(\frac{\varepsilon}{k_B T} \right )-1  \right ]^{-1}$ is the Fermi-Dirac distribution.
Note that, according to this definition, the relative thermal conductance is dimensionless
and express clearly the ratio between the actual conductance and the maximum achievable  thermal conductance  for a quantum channel which is $G_T$.
The transmission function, ${\cal T}(\varepsilon)$, can be written in terms of the probability scattering coefficients as
\begin{equation}\label{eq:transmission}
{\cal T}(\varepsilon) = \sum_{\alpha,\beta = \tilde{e},\tilde{h}} P^{\alpha,\beta}_{R,L},
\end{equation}
with $P^{\alpha,\beta}_{R,L}$ given by Eq. (\ref{eq: Pmatrix}).
We recall that in the next section, for sake of simplicity, we will focus on $l_S =0$. Similar results are obtained when $l_S \neq 0$.

\section{Results}\label{sec:results}
We now turn to discuss results for the Andreev spectrum obtained by solving Eq. (\ref{eq: compatibility}) for two configurations: the single-domain magnetic island  (Sec. \ref{sec:one-island-abs}) and the two-domain island (Sec. \ref{sec: two islands-abs}) in the Josephson junction.
The properties  of the Andreev states strongly affect the behavior of the dc Josephson current. Although the quasiparticle states above the gap also contribute to the Josephson current, the signatures of the topological phase, like the jumps in the current-phase relation in the dc case and the periodicity in the ac case, fully depend on the behavior of the Andreev states \cite{fu-kane,houzet,jiang,laurens,marra2016}. Instead, the latter do not play any direct role in the response to the difference of temperature $\delta T$ between the two superconductors. 
The latter manifests itself in the thermal conductance, which we analyze  for both configurations in Sec. \ref{sec:one-island-tau} and Sec. \ref{sec.two-island-tau}. 
As already mentioned, we focus on configurations with $l_S=0$, since this parameter does not affect the main results we aim to discuss. In addition, we find sometimes convenient to characterize 
the strength of the magnetic coupling with respect the proximized superconducting gap $\Delta_0$ through the
dimensionless parameter 
\begin{equation}
    \Gamma = \frac{Jm}{\Delta_0} = \frac{\xi_S}{\xi_M}.
\end{equation}
 Importantly, as highlighted in the last equality, this parameter also defines the ratio
between the magnetic and superconducting lengths.

\subsection{Andreev spectrum of a junction with a single-domain magnetic island}\label{sec:one-island-abs}
The Andreev bound states for the configuration corresponding to a magnetic island with a  single magnetic domain  have been already analyzed in Refs. \onlinecite{fu-kane,houzet,crepin}. Here, we review  those results in order to have them as a reference. 

\begin{figure}[htp]
	\centering
	\includegraphics[width=\columnwidth]{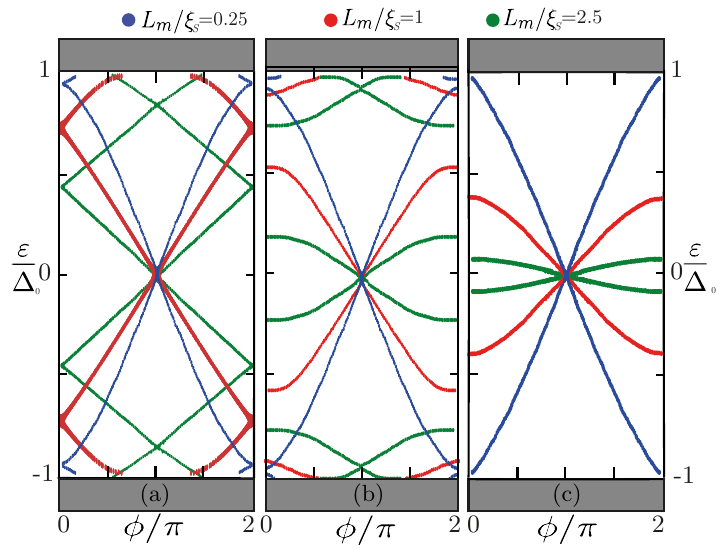}
	\caption{Andreev spectrum of the junction with a single magnetic domain of different lengths. Panel (a), (b) and (c) correspond to $\Gamma =0,~0.5,~ 1$, respectively. Grey bands indicate the continuum spectrum. Different colors correspond to different lengths $L_m/\xi_S = 0.25, 1, 2.5$ of the magnetic island. }
	\label{fig:ABS-F}
\end{figure}

Figure \ref{fig:ABS-F} presents the Andreev spectra calculated  for several lengths $L_m$ of the magnetic island. 
The left panel of Fig. \ref{fig:ABS-F} shows the spectrum for $\Gamma =0$, which corresponds to a junction hosting bare helical edge states between the superconducting contacts. We can identify two degenerate states corresponding to a Kramers pair at the time-reversal symmetric case $\phi =0, \mathrm{mod}(2\pi) $. The degeneracy is broken as $\phi$ advances with one of the states evolving to a higher energy and hybridizing with the quasiparticle continuum for $|\varepsilon| > \Delta$. A crossing point at zero energy takes place at $\phi = \pi$. For $\Gamma \neq 0$ time reversal symmetry is broken even for $\phi=0$, and the degeneracy is consequently lifted.
According to calculations \cite{fu-kane,crepin}, Majorana modes are stabilized at the boundaries of the magnetic domain in the present case. Hence, the Andreev states with lowest absolute value of the energy result from the hybridization of these Majorana modes. These two states have different parity and cross at $\phi = \pi$. Since they are completely decoupled from the quasiparticle continuum, the spectrum is effectively $4\pi$-periodic and so does the ac Josephson current if parity is conserved, in contrast to the $\Gamma=0$ case, which is $2\pi$-periodic due to the hybridization of the subgap states  with the continuum.
 An interesting feature to highlight is the fact that 
the Andreev spectra are qualitatively different in the case $\Gamma<1$ (panel (b) of Fig \ref{fig:ABS-F}) and $\Gamma\geq1$ (panel (c) of Fig. \ref{fig:ABS-F}). For $\Gamma <1$, several Andreev bound states may exist in the gap for large enough junctions, in addition to the ones with lowest absolute energy. Instead, for $\Gamma \geq 1$ the spectrum has only two 
 Andreev states, which result from the hibridization of the two 
 MZMs.

\subsection{Andreev spectrum of a junction with a magnetic island with two domains} \label{sec: two islands-abs}
\begin{figure}[htp]
	\centering
	\includegraphics[width=\columnwidth]{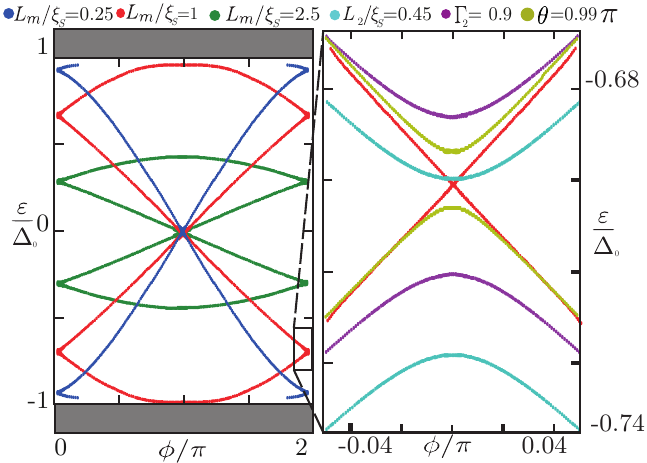}
	\caption{Andreev spectrum of the junction with  two magnetic domains. Left panel correspond to $\Gamma=1$ and $\theta=\pi$ and each domain of length $L_1=L_2=L_m/2$. Different colors correspond to different lengths, $L_m/\xi_S = 0.25, 1, 2.5$ of the full magnetic island. Grey band indicates the continuum spectrum. Black box indicates zoom on spectrum around $\phi=0,\mathrm{mod}(2\pi)$ for two domains. Right panel: Andreev spectra for two domains, in the range covered by the box of the left panel. 
	Different lines correspond to a single parameter variation: $L_2 = 0.45\xi_S$ (light blue), $\theta=0.99\pi$ (yellow) and $\Gamma_2 = 0.9$ (violet). Red line is the same in both panels.
	}
	\label{fig:ABS-AF}
\end{figure}

The spectrum for two domains of lengths 
$L_{1,2} = L_m/2$,
equal magnetizations 
$\Gamma_{1,2}= \Gamma = Jm/\Delta$ 
and relative tilt $\theta = \pi $ in the orientation of the magnetization of the domains is presented in the left panel of Fig \ref{fig:ABS-AF}. 

For $\phi =0$, this system is invariant under spatial inversion symmetry with respect to 
the center of the junction, i.e. $x=x_S/2$
and the simultaneous inversion of the magnetic moments and spin. For this reason, 
in this configuration, the Andreev bound states are degenerate for $\phi =0, \mbox{mod}(2\pi)$ (see the level crossing in the left panel of Fig. \ref{fig:ABS-AF}). In general this degeneracy is broken and a gap appears in the spectrum.
A representative example is illustrated in the right panel of Fig \ref{fig:ABS-AF}. This panel shows a zoom at  $\phi =0,\mathrm{mod}(2\pi)$ for 
configurations which slightly depart from the symmetric case with $L_m/\xi_S =1$ (red curve in both panels of Fig. \ref{fig:ABS-AF}). Namely, for the light blue curve we set $L_2 = 0.45\xi_S$, for the yellow curve $\theta = 0.99 \pi$ and for the violet we set $\Gamma_2 = 0.9 $ while all the other parameters are equal to the symmetric case. The plot in red lines is a reference equal to the red one in the left panel.
For all the configurations examined, the crossing at $\phi = \pi$ is 
topologically protected,
as in the case of a single magnetic domain. 
The behavior as a function of the coupling $\Gamma$ is also similar to the case of a single magnetic domain analyzed in Fig. \ref{fig:ABS-F}. Namely, for $\Gamma>1$ the spectrum is composed of only two Andreev states crossing at $\phi=\pi$, which can be identified as hybridized Majorana states. 

\subsection{Thermal conductance of a junction with a  single-domain magnetic island}\label{sec:one-island-tau}
\begin{figure}[htp]
	\centering
	\includegraphics[width=\columnwidth]{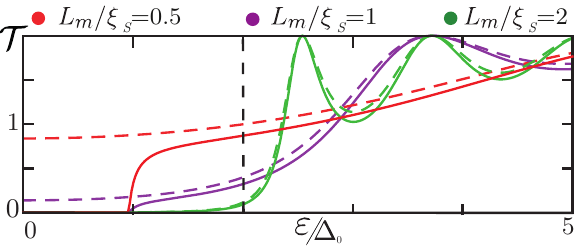}
	\caption{Transmission function for a single magnetic domain of length $L_m$ embedded in the Josephson junction with $\Gamma =2$ ($\xi_M=\xi_S/2$) and $\phi=0$. Solid lines correspond to $T=0.44 T_C$ while dashed lines correspond to $T= 1.1 T_C$. Vertical line indicates the magnetic gap. Other details are in the figure.
	}
	\label{fig:Tau-F}
\end{figure}

The relative thermal conductance, defined in Eq. (\ref{eq:kappa}) is completely determined by the behavior of the transmission function given in Eq. (\ref{eq:transmission}). The analytical expression 
of the transmission function for a system with a magnet and
without superconductors has been presented in Refs.~\onlinecite{optimal,Bustos,dolcini}. 
Here, instead, we discuss the numerical results of the transmission function introduced in Eq. (\ref{eq:transmission}) for the hybrid system with the superconducting contacts. In this configuration,
in addition to the magnetic gap, the transmission function  depends on the effect of the superconducting gap which has a strong temperature dependence.
We approximated the usual self-consistence dependence of the BCS theory for the superconducting gap as a function of the temperature, with $\Delta(T) = \Delta_0 \tanh\left(1.74\sqrt{\frac{T_C}{T}-1}\right)$ being $\Delta_0$ the corresponding value at $T=0$. Since in the present problem $\Delta_0$ is the gap induced by proximity effect on the 2DTI, it is expected to be smaller than the corresponding value  in the bulk of the superconducting contact. On the other hand, the magnetic gap does not change in magnitude within the temperature range 
we will consider hereafter.
Therefore,  the transmission function depends on the temperature $T$, as well as on the
amplitude of the magnetic coupling,
governed by the dimensionless parameter, $\Gamma$ 
and the length of the magnet $L_m$. This is illustrated in Fig. \ref{fig:Tau-F}, where results for the transmission function are presented for the dimensionless magnetic coupling $\Gamma=2$ and several lengths of the magnetic island at temperatures $T<T_C$ (solid lines) and $T>T_C$ (dashed lines).  Notice that for 
$T>T_C$ the superconducting gap is closed. Hence, in this case, the transmission function coincides with the one for a magnetic island contacting the helical edge states without superconductivity analyzed in Ref. \onlinecite{optimal}.

Focusing on the plots 
of Fig. \ref{fig:Tau-F} we see that 
the quasiparticle
transmission is strongly suppressed for the case $L_m = 2\xi_S$ (corresponding to $L_m=4 \xi_M$) up to energies $\varepsilon \simeq 2 \Delta_0$, for both temperatures. This is a clear manifestation of the fact that a gap opens in the spectrum
of the edge states when the length of the island is $L_m\gg\xi_M$, while for $T>T_C$ the gap is not fully developed for islands of length $L_m \simeq \xi_M$ or shorter. 
When the magnetic gap is not fully formed the superconductivity can suppress the transmission. Indeed,
within the low temperature regime shown in solid lines, superconductivity dominates, and the superconducting gap erases all the spectral features with energies $\varepsilon<\Delta(T)$. This implies that for the short magnets considered in the figure, (with length $L_m<\xi_M$) the spectrum is still gapped, while there is a finite spectral weight at higher temperatures when the superconductivity is suppressed. 
On the other hand at energies $\varepsilon>\Delta(T)$, the transmission function has a structure of peaks and minima that depends on $L_m$. For low temperatures, the features above $\Delta(T)$ also depend on the superconducting phase difference $\phi$ (that was set to zero in the Fig. \ref{fig:Tau-F}).
In conclusion,
for the long islands, like the one shown in the figure with $L_m = 2\xi_S=4 \xi_M$, we can clearly see the dominance of the magnetic gap over the superconducting one. In fact, there is a gap of $~\Gamma\Delta_0$ (see vertical dashed line) in the transmission function in both regimes of temperatures, in strong contrast with the cases of the shorter magnets, with  $L_m = \xi_M, ~ 2 \xi_M$. This will be reflected in the behavior of the thermal conductance, to be discussed shortly.

\begin{figure}[htb]
	\centering
	\includegraphics[width=\columnwidth]{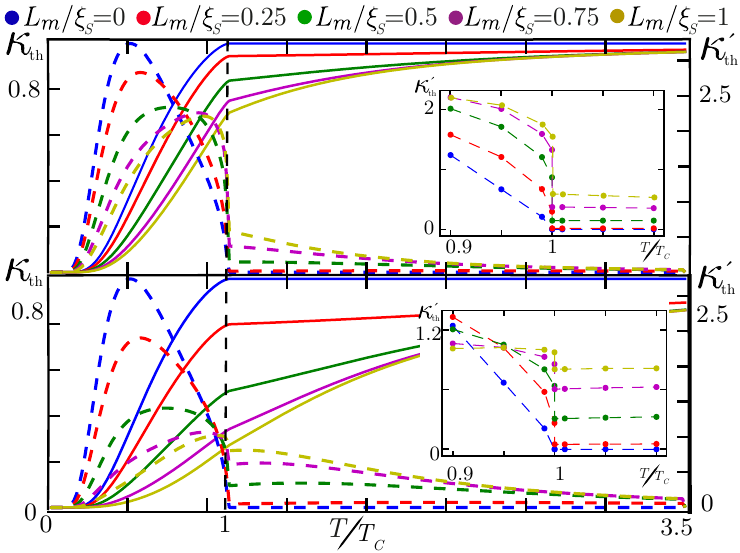}
	\caption{
	Relative thermal conductance $\kappa_{\mathrm{th}}(T)$ (left axis, solid line) and its temperature derivative $\frac{d \kappa_{\mathrm{th}}}{d T} = \kappa_{\mathrm{th}}'$(right axis, dashed line)
	for a magnet with a single magnetic domain embedded in the Josephson junction with $\phi=0$. 
	Top panel corresponds to $\Gamma =1$ and bottom panel corresponds to $\Gamma =2$. Black dashed vertical line indicates $T= T_C$. Inset: zoom of $\kappa'$ near $T = T_C$ .
	Other details are on the figure.
	}
	\label{fig:KappaFerro}
\end{figure}

After the analysis of the transmission function we discuss the relative  thermal conductance, $\kappa_{\rm th}$.
Indeed, the competition between the magnetic and the superconducting gaps is particularly evident in the behavior of this response function as a function of temperature. 
This is shown in Fig. \ref{fig:KappaFerro}.
The upper panel corresponds to $\Gamma=1$ (where $\xi_M = \xi_S$) and the lower one to $\Gamma =2$ (where $\xi_M = \xi_S$/2). 
In all the plots we can clearly identify the exponentially small value of the thermal conductance at low temperatures as well as the high temperature saturation to the quantum bound $\kappa_{\rm th}=1$ \cite{Pendry-Limit,Bekenstein1,Bekenstein2}
for $L_m\to 0$. The limit  where $L_m=0$ corresponds to the junction without magnetic island, in which case, the transport channel is fully open only when the superconducting gap closes, for $T>T_C$.

At finite $L_m$, the magnetic gap remains open even when the superconducting gap is closed, and the thermal conductance 
is smaller than the quantum bound.
This feature is enhanced with increasing $\Gamma$ (compare both panels of the figure). The changes in  $\kappa_{\rm th}$ as the superconducting gap closes are more visible in the behavior of its 
derivative $\kappa'_{\rm th} = d \kappa_{\rm th}/dT$, which is shown in each panel with dashed lines. At $T=T_C$ the derivative has a discontinuity, as expected when the phase transition happens between a superconducting regime and a non-superconducting one. For $T>T_C$ the derivative of the thermal conductance monotonically increases with both $\Gamma$ and $L_m$, while the opposite behavior takes place for $T<T_C$.

\begin{figure}[htp]
	\centering
	\includegraphics[width=\columnwidth]{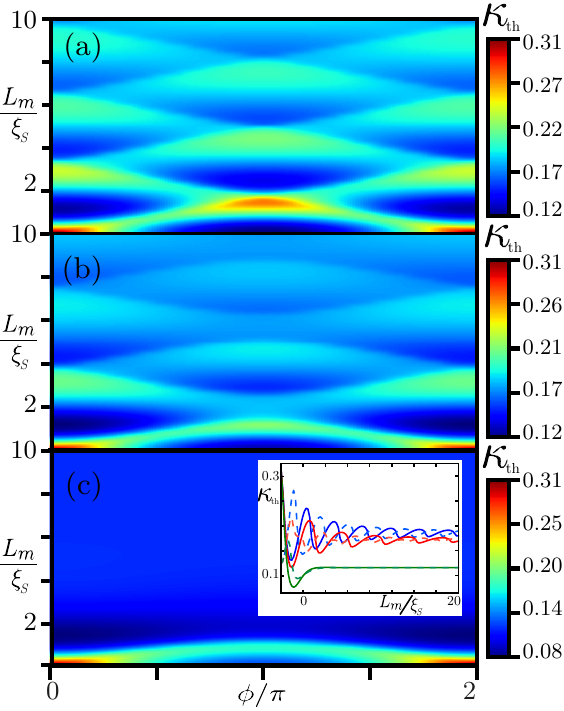}
	\caption{
	Relative thermal conductance of the junction with a single magnetic domain with temperature $ T = 0.44 T_C$.
	Different panels correspond to different ratios $\Gamma=Jm/\Delta_0$. Panel (a) corresponds to $\Gamma =0$, panel (b) to $\Gamma = 0.5$ and panel (c) to $\Gamma = 1$. Inset in panel (c) is $\kappa_{\mathrm{th}}$ as function of $L_m/\xi_S$, blue line corresponds to $\Gamma =0$, red line corresponds to $\Gamma = 0.5$ and green line corresponds to $\Gamma = 1$. Solid lines are for $\phi =0$ and dashed lines are for $\phi = \pi$.
	}
	\label{fig:Colormap-F}
\end{figure}

When $T>T_C$ there is no superconductivity, hence 
it cannot be defined a fixed 
phase bias $\phi$ in the setup. On the other hand, 
when $T<T_C$ there exists a phase difference $\phi$ in the Josephson junction which introduces quantum interference in  the behavior of the thermal conductance. We better analyze the features related to this effect in  Fig. \ref{fig:Colormap-F}, which
shows the behavior of $\kappa_{\rm th}$ as function of $\phi$ and $L_m$ for a fixed  temperature $T<T_C$ and different values of $\Gamma$ in the different panels. The limit $\Gamma=0$, corresponding to the junction without magnetic island, has been previously analyzed in Refs. \onlinecite{soth-han2,nonloc2} and is shown in panel (a).
The main feature to highlight is the oscillatory response 
which is even and $2\pi-$ periodic in $\phi$
and oscillatory but decreasing on $L_m/\xi_S$, as shown in panel (b) for $\Gamma=0.5$. Further details on the oscillatory behavior as a function of the length are presented in the inset of the panel (c), where results for $\kappa_{\rm th}$ 
as function of $L_m$ are shown only for $\phi=0,~\pi$ and $\Gamma =0,0.5,1$ up to length $L_m=20\xi_S$.  We see that, for finite $\Gamma<1$, the pattern of damped oscillations is very similar to the one without magnet (corresponding to $\Gamma=0$). Notice, in particular that, besides a shift and a smaller amplitude, the period of the oscillations is basically the same in the cases with $\Gamma= 0,~0.5$.
Albeit, as the strength of the magnetic coupling is increased and overcomes $\Gamma=1$, this response is much less sensitive to $\phi$ and decreases very fast with the length. This is consistent with a behavior dominated by the magnetic gap, even for temperatures below $T_C$, where the superconducting gap is finite. 
In the inset of panel (c) it  can be appreciated  how the thermal conductance tends to some limit when $L_m\gg \xi_S$ that depends on $\Gamma$ but does not depend on $\phi$. This saturation value is achieved as a limit of the damped oscillations for $\Gamma <1$, while it is approached fast and without oscillations for $\Gamma \geq 1$.

\subsection{Thermal conductance for a magnetic island with two magnetic domains}\label{sec.two-island-tau}

\begin{figure}[htp]
	\centering
	\includegraphics[width=\columnwidth]{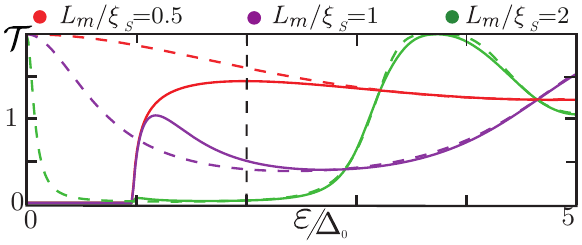}
	\caption{Transmission function for two magnetic domains of length $L_1=L_2=L_m/2$  with opposite orientation, $\theta=\pi$, and equal magnetization $m$,  embedded in the Josephson junction for $\Gamma=2$ and $\phi=0$. Solid lines correspond to $T=0.44 T_C$ while dashed lines correspond to $T=1.1 T_C$. Vertical dashed line indicates the magnetic gap. 
	Other details are on the figure.
	}
	\label{fig:TauAF}
\end{figure}

The transmission function for the magnet with two domains is shown in Fig. \ref{fig:TauAF}
for $T<T_C$ (solid lines) and $T>T_C$ (dashed lines), for $\theta=\pi$ and $\Gamma=2$ with $L_{1,2}=L_{m}/2$. In the absence of superconducting contacts, or equivalently, for $T>T_C$, the main feature  is the presence of resonant peaks inside the gap, as it was discussed in Ref \onlinecite{optimal}. 
In  this case we can clearly distinguish 
the resonance that develops at $\varepsilon=0$. 
This peak has a width that decreases with the length of the island $L_m$ and for these parameters it corresponds to a JR zero mode. For other relative orientations $\theta \neq \pi$ the resonance is shifted from $\varepsilon=0$, albeit remains being a robust feature within the gap for a wide range of parameters \cite{optimal,goldstone,hughes,flecks}.
The value of the expected magnetic gap, $\Gamma \Delta_0$,
corresponding to the island with uniform magnetization is indicated as a reference with a vertical line in the figure. 

\begin{figure}[htp]
	\centering
	\includegraphics[width=\columnwidth]{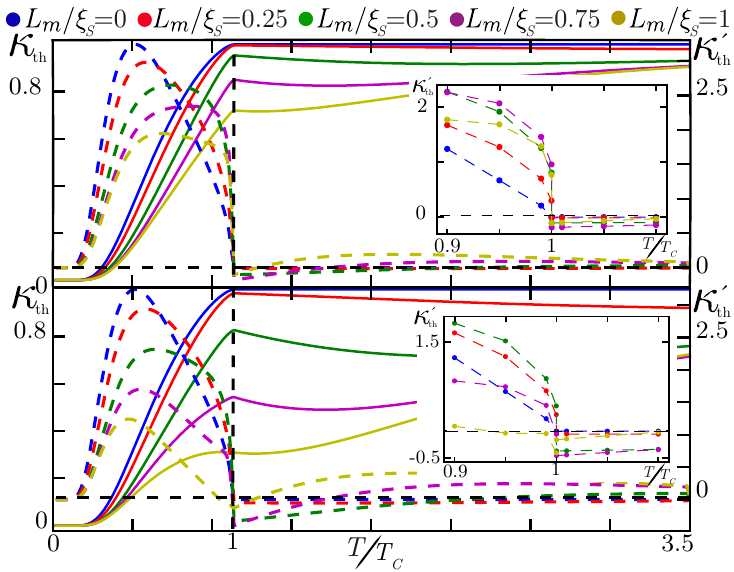}
	\caption{
	Relative Thermal conductance $\kappa_{\mathrm{th}}(T)$ (left axis, solid lines) and its derivative $\kappa'_{\mathrm{th}}$ (right axis, dashed line) for two magnetic domains with 
	opposite orientations, equal length $L_m/2$ and equal magnetizations $m_1 =m_2 =m$ for $\phi=0$.
	Top panel corresponds to $\Gamma = 1$ and bottom panel corresponds to $\Gamma =2$. Black dashed horizontal line on both panels indicates $\kappa'_{\rm {th}}=0$ while vertical line indicates $T = T_C$.
	Inset shows a zoom of $\kappa'$ near $T = T_C$.
	Other details are on the figure.
	}
	\label{fig:kappa-af}
\end{figure}
In the plots corresponding to
$T<T_C$ we can clearly see the effect of the superconducting gap,
i.e. the transmission function is vanishing for $\varepsilon<\Delta(T)$.
Anyway the JR resonance in the gap leads to a remarkable  behavior of the thermal conductance as a function of the temperature, which 
 is presented in Fig. \ref{fig:kappa-af} for a configuration with two domains with equal length, $L_{1,2}=L_m/2$, equal magnetization  and opposite orientation, i.e. $\theta=\pi$.
As in the case of the island with a single domain,  when $L_m\to 0$, $\kappa_{\rm th}$ tends to saturate at the quantum bound for $T>T_C$ and is exponentially small at low temperatures.
However, we now see that the derivative $\kappa'_{\rm th}$ is negative right above $T_C$. 
We can trace back this peculiar feature to the development of the resonant peak in the gap as the temperature overcomes the critical temperature.  From the mathematical point of view, this can be understood by calculating the derivative with respect to the temperature on Eq. (\ref{eq:kappa}), which leads to
\begin{equation}
    \kappa_{\rm th}' = \frac{2\kappa_{\rm th}}{T}-\frac{1}{G_T T}\int_0^\infty \varepsilon^3
\frac{\partial f}{\partial \varepsilon}\frac{\partial {\cal T}}{\partial \varepsilon} d\varepsilon, ~~~~~ T\geq T_C.
\end{equation}
The first term on RHS is due to the contribution of  $G_T$ -- recall that this quantity is linear with $T$ -- and is always positive. Instead, the sign of the second term depends on the sign of the derivative of the transmission function. Therefore, since $\partial f/\partial \varepsilon <0$, if $ \partial {\cal T}/\partial \varepsilon $ is negative and the contribution of the second term is large enough, the derivative of the relative thermal conductance may be negative. This is precisely the case of 
the configuration with two magnetic domains due to the resonance where,  within the window defined by the function $-\partial f/\partial \varepsilon$, the transmission function ${\cal T}(\varepsilon)$ has a negative slope, which leads to a large contribution to the integral when multiplied by $\varepsilon^3$. This contribution becomes small as the length of the island increases and the resonance becomes narrow enough. In conclusion, the result of having $\kappa'_{\rm {th}}$ negative just above $T_C$ can be regarded as an indication of the presence of a JR peak in a Josephson junction.
\begin{figure}[htp]
    \centering
    \includegraphics[width =\columnwidth]{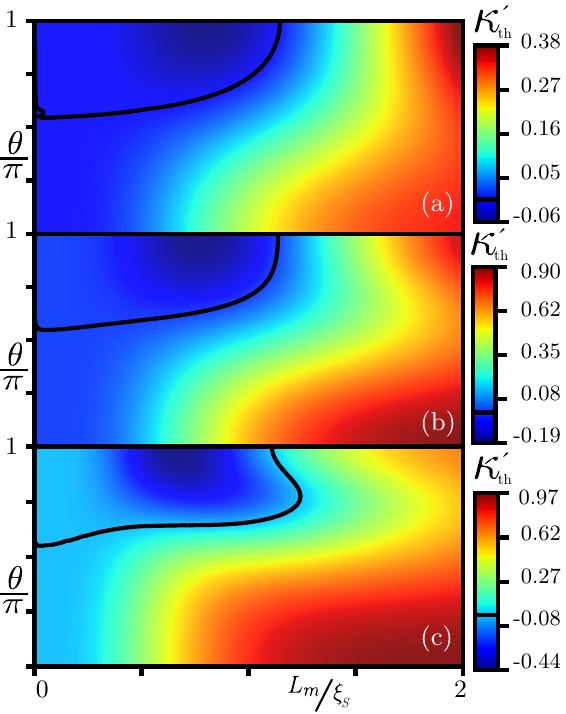}
    \caption{Derivative of the relative thermal conductance, $\kappa'_{\mathrm {th}}$ at $T = 1.01 T_C$ as function of the relative tilt $\theta$ in the orientation of the magnetic moments with $L_m/2 = L_{1} = L_{2} $ and $\phi=0$. Panels (a), (b) and (c) correspond to $\Gamma = 0.5, 1, 2$, (with $\xi_M=2\xi_S, \xi_S, \xi_S/2$) respectively.  The black line in each panel indicates the boundary for the region with $\kappa'_{\mathrm{th}}<0$.}
    \label{fig:DkappaMin}
\end{figure}

This effect is analyzed in more detail in Fig. \ref{fig:DkappaMin}, where $\kappa'_{\rm th}$ is shown as function of the relative orientation of the islands and the length for fixed $\Gamma$ in each panel at a temperature just above $T_C$. We can see that there is a wide range of lengths and orientations close to $\theta = \pi$ where
$\kappa'_{\rm th}$ is negative. These cases coincide with configurations leading to resonant peaks of the transmission function inside the magnetic gap. Importantly, the width of the resonant peak scales with the inverse of the length of the magnetic island (see plots in dashed lines in Fig. \ref{fig:TauAF}). Hence, the impact of this feature in generating a negative derivative of the relative thermal conductance just above $T_{C}$ becomes negligible as the length of the magnetic island increases.
This shows that the JR peak can be identified by the negativity of $\kappa'_{\rm th}$ only if $L_m\leq\xi_S$ i.e. for sufficiently short magnetic islands.
\begin{figure}[htp]
	\centering
	\includegraphics[width=\columnwidth]{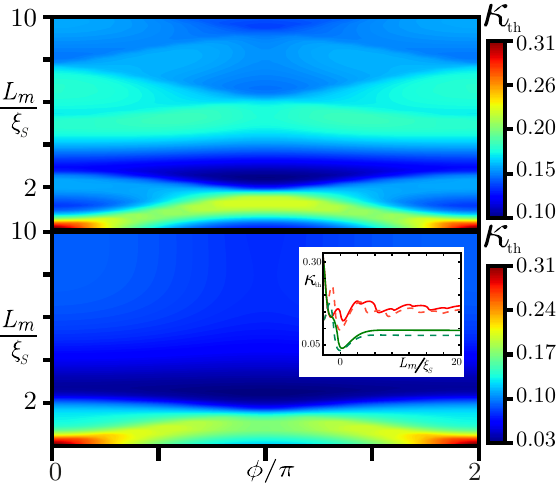}
	\caption{Relative thermal conductance of the junction with two magnetic domain with opposite magnetic moments and $L_{1} = L_{2} = L_m/2 $ at temperature $T = 0.44 T_C$. Upper and lower panels correspond to  $\Gamma = J_m/\Delta_0=0.5,~1$ ($\xi_M=2\xi_S,~\xi_S$). The inset in the lower panel shows $\kappa_{\mathrm{th}}$ as function of $L_m/\xi_S$. Red line is for $\Gamma=0.5$ and green line is $\Gamma=1$ for $\phi = 0, ~(\pi)$ for  solid (dashed) lines. }
	\label{fig:ColormapAF}
\end{figure}

As in the case of a single-domain configuration, we expect some dependence of the thermal conductance on $\phi$ in the low-temperature regime with $T<T_C$. This is analyzed in
Fig. \ref{fig:ColormapAF} for the case of two magnetic domains with opposite orientations of the magnetic moments ($\theta=\pi$). The interference pattern is still even and $2\pi-$ periodic in $\phi$ but different from the one observed in Fig. \ref{fig:Colormap-F} for a single domain. However, like in that case, as the length of the magnetic island increases and becomes significantly larger than $\xi_M$, the magnetic gaps becomes dominant, and the features introduced by $\phi$ become suppressed. This is highlighted in the inset shown in the bottom panel of the figure and we can pose similar observations as in the case of the single magnetic domain. Namely, for $\Gamma \geq 1$, where the magnetic gap dominates, the conductance is practically non-sensitive to the superconducting phase $\phi$ and it decreases 
rapidly with $L_m$. Instead, for $\Gamma<1$, thermal conductance depends on $\phi$ and displays oscillations as a function of $L_m$. The pattern of such oscillations is very different and much less regular than the one observed in the junction without magnetic island (corresponding to $\Gamma=0$). Hence, in the regime of $\Gamma <1$, the interference pattern of the thermal conductance provides clear signatures of the domain structure of the magnetic island. Instead, for $\Gamma \geq 1$, the  rapid suppression of the thermal conductance is an indication of the effect of the magnetic island, but no information on the domain structure can be extracted from that behavior.

\begin{figure}[htp]
	\centering
	\includegraphics[width=\columnwidth]{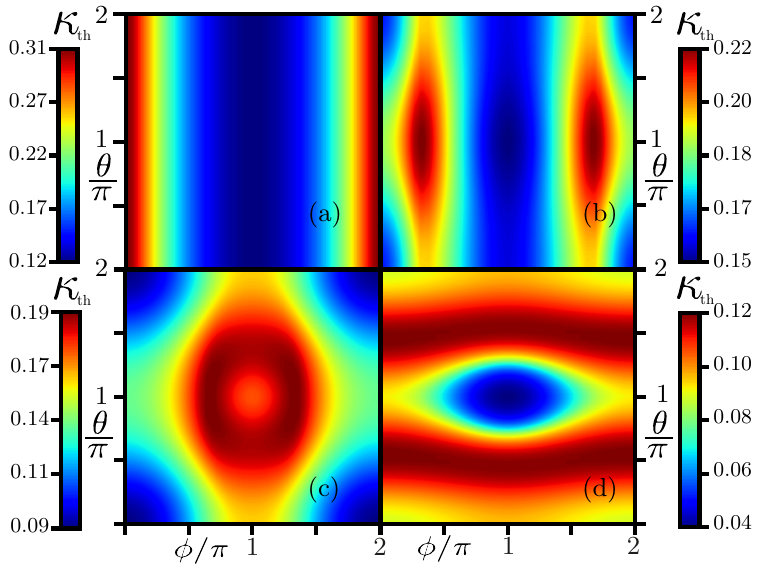}
	\caption{
	Relative thermal conductance $\kappa_{\mathrm{th}}(T=0.44T_C)$ as function of the tilting angle in the orientation of the magnetic domains  $\theta $ and the phase difference of the superconducting potentials $\phi$ for $\Gamma =1$ ($\xi_M=\xi_S$) and $L_{1} = L_{2}=L_m/2$. Panel (a) corresponds to $L_m/\xi_S = 0$, panel (b) to $L_m/\xi_S = 0.5$, panel (c) to $L_m/\xi_S = 1$ and panel (d) to $L_m/\xi_S = 2$.
	}
	\label{fig:theta-phi}
\end{figure}

Finally in Fig. \ref{fig:theta-phi}, 
we analyze the combined effect of $\theta$ and $\phi$ 
on the behavior of the thermal conductance for different lengths of the magnetic island, a magnetic coupling corresponding to $\xi_M =\xi_S$ where we expect the maximal interplay between superconductivity and magnetic scales, and setting $T<T_C$.
The figure highlights the fact that, not only the superconducting phase bias $\phi$ generates interference patterns but also the tilting angle $\theta$.
Furthermore, we notice that the specific features as function of $\phi$ are similar to those as function of $\theta$. This is not surprising in the view of the duality relation between these two parameters\cite{jiang}. 
Notice that such duality implies that similar physical properties should be observed if the magnetic islands are interchanged with the superconductors, with $\theta$ playing the role of $\phi$ and viceversa, due the similar structure of the massive terms, as is explicit in Eq. (\ref{hmdel}). In the Josephson-junction configuration studied here, we can observe signatures of the aforementioned duality, in a context where
the superconductors have infinite length, while the magnetic islands are finite.
In fact, we see that panel (a), which corresponds to a purely superconducting junction has a pattern of straight vertical features, reflecting the sensitivity of the thermal conductance only with the phase bias $\phi$. In the opposite limit of a long enough magnetic island shown in panel (d), the magnetic effect becomes dominant and the pattern tends to follow horizontal straight lines, indicating a sensitivity on the tilt $\theta$ but loosing the dependence on $\phi$. Configurations between these two cases can be observed in panels (b) and (c). These results show that the interplay of the tilting angle and phase difference may be an interesting phenomenology in the studied system.

\section{Conclusions}\label{sec:con}
We have analyzed the Andreev spectrum and the thermal conductance of a one edge Josephson junction of a 2D topological insulator hosting a magnetic island with one and two magnetic domains. We have shown that the Andreev spectrum, which defines the behavior of both the dc and ac Josephson current, is qualitative similar for these two configurations of islands.
Instead, the behavior of the thermal conductance shows several features as a function of $\phi$ and the temperature $T$ that characterize the nature of the junction. We have analyzed in detail all these properties. Most of them can be understood as a consequence of a competition between the temperature-dependent superconducting gap $\Delta(T)$ below the superconducting critical temperature $T_C$ and the magnetic gap, which typically remains constant within this temperature range. 

A remarkable result is the fact that for configurations with two magnetic domains with different orientations, which host JR resonant modes, the thermal conductance decreases with the temperature, just above the superconducting critical temperature. This is a peculiar behavior that could be useful to identify the existence of these intriguing modes. 
So far, no experimental signatures of JR resonances have been reported and this signature in the thermal conductance can be useful to identify them.
Notice that the Josephson current is not sensitive to the existence of this resonant state.
The Josephson current is defined by the derivative of all the negative energies of the spectrum of the superconducting junction with respect to the phase bias, including the quasiparticle continuum and the subgap Andreev states. We have shown that the spectra for systems where the magnets have different orientations are similar. Furthermore, they are also similar to those without any magnet. These results rule out the Josephson current as an appropriate witness of the existence of a JR mode.
More importantly, the peak in the transmission probability associated to the JR resonance develops when the superconducting gap closes, in which case there is no Josephson effect at all.
 Instead, the thermal conductance, being a non-equilibrium quantity, depends not only on the spectrum but also on the transmission properties of the system, irrespectively of the fact that the system is superconducting or not. The development of the JR resonance above $T_C$ generates a large transmission probability at low energies, which significantly affects the behavior of the thermal conductance.  

In the low-temperature regime, for $T<T_C$, 
the thermal conductance in the two-domain configuration shows 
 interference patterns  as a function of both the phase bias of the superconductors and
the angle between the magnetic moments. This feature is particularly clear for islands where the magnetic and superconducting lengths are similar, $\xi_M \sim \xi_S$. 

According to estimates presented in Ref. \onlinecite{optimal}, reasonable configurations of the magnetic island, compatible with the present state of the art experiments, should have magnetic lengths below  $\xi_M \sim 10-20 \mu m$, which is of the same order of magnitude of the superconducting coherence length $\xi_S$ and energy gaps of $Jm \sim 1.2 - 2.4 K$. These  correspond to a regime with $\Gamma \sim 1,2$, similar to the one analyzed in the present work, where the different features of $\kappa_{\rm th}$ as a function of $\phi$ below $T_C$, as well as the corresponding behavior as a function of temperature close to $T_C$, clearly
distinguish the different type junctions. 

\section{Acknowledgements}
We acknowledge support from CNR-CONICET cooperation program ``Energy conversion in quantum nanoscale hybrid devices''.  We are sponsored by PIP-RD 20141216-4905 
of CONICET,   PICT-2017-2726 and PICT-2018-04536 from Argentina, as well as the Alexander von Humboldt Foundation, Germany, the ICTP-Trieste through a Simons associate program (L. A.) and the Federation Institute (D. G.). G. M., A. B. and F. T. acknowledge SNS-WIS joint lab QUANTRA. M. C. is supported by the Quant-Era project ``Supertop''. A. B. acknowledge the Royal Society through the International Exchanges between the UK and Italy (Grant No. IEC R2 19216)
\appendix
\section{Scattering matrix}\label{sec: Appendix}

In this appendix we present the details of the calculations for the different scattering matrices $S_{L,R}$ and $S_M$ introduced in Eq. (\ref{eq: sm combinated main}) of the main text. We proceed by writing first the scattering matrices $S_{L,R}$ describing the left and right interfaces of the 2DTI with the superconductors (in Sec. \ref{sms}) and then the scattering matrix $S_M$ describing the 2DTI edge in contract with the magnetic domain (in Sec. \ref{smm}). In section \ref{comb} we combine them in order to get the full scattering matrix of the system.

\subsection{Scattering matrix of the SC-2DTI interface}\label{sms}
The scattering matrix equation for $S_L$, obtained by solving the wave function matching problem at the interface between the left superconducting lead and the edge state of the 2DTI, can be written as
\begin{equation} \label{eq: scattering matrix-sc}
   \begin{pmatrix}
\tilde c_{L}^{-}\\ 
\tilde b_{L}^{-}\\
c_{L}^{+}\\  
b^{+}_{L}
\end{pmatrix}=
\begin{pmatrix}
0 &r^L_{\tilde e,\tilde h}& t^L_{\tilde e,e}   &0 \\ 
r^L_{\tilde h,\tilde e} &0  &0  &t^L_{\tilde h,h} \\ 
t^L_{e,\tilde e} &0  &0  &r^L_{e,h} \\ 
0 &t^L_{h,\tilde h} & r^L_{h,e}   &0 
\end{pmatrix}
\begin{pmatrix}
\tilde c_{L}^{+}\\ 
\tilde b_{L}^{+}\\ 
c_{L}^{-}\\ 
b^{-}_{L}
\end{pmatrix},
\end{equation}
where we indicated with $c^{\mp}_{L}/b^{\mp}_{L}$ the incoming and outgoing quasiparticles/quasiholes in the TI region and with $\tilde c^{\pm}_{L}/\tilde b^{\pm}_{L}$ the incoming and outgoing quasiparticles/quasiholes in the superconductor. The index $L$ labels the interface with left superconductor and $\pm$ indicate the direction of propagation of quasiparticles along the x-axis ($+$ for right movers and $-$ for left movers). 
Notice that with this basis the scattering matrix can be written as 
\begin{equation}
\label{eq: SL}
S_L = \begin{pmatrix}
r_L & t'_L \\ 
t_L & r'_L 
\end{pmatrix},    
\end{equation}
where $r_L$ and $r_L^{\prime}$ are blocks concerning particles reflected at the interface, whereas $t_L$ and $t_L^{\prime}$ are blocks concerning particles transmitted through the interface.
The obtained coefficients $r^L_{\alpha,\beta}$ and $t^L_{\alpha,\beta}$ represent the reflection and transmission amplitudes respectively of an incoming particle of type $\beta $ to a particle of type $\alpha$ at the interface. 

The coefficients of Eq. \ref{eq: scattering matrix-sc} can be compactly written as
\begin{align} \label{eq: coefficients}
r^L_{\gamma,\bar \gamma} &= \gamma \frac{v}{u}e^{i\alpha}e^{i\frac{\phi}{2}}, \nonumber\\ 
r^L_{\tilde \gamma,\tilde{\bar \gamma}} &=-\frac{v}{u}e^{-i\beta}\Theta\left(\varepsilon-\Delta \right ) ,\nonumber\\ 
t^L_{\gamma,\tilde\gamma} &= \frac{\sqrt{u^2-v^2}}{u}e^{\frac{i}{2}(\alpha-\beta)}e^{-i\gamma \frac{\phi}{4}} \Theta(\varepsilon-\Delta), \nonumber \\ 
t^L_{\tilde \gamma,\gamma} &= \bar \gamma \frac{\sqrt{u^2-v^2}}{u} e^{\frac{i}{2}(\alpha-\beta)}e^{i\bar \gamma \frac{\phi}{4}} \Theta(\varepsilon-\Delta),
\end{align}
where the QP/QH index $(\gamma = e,h)$ in the LHS is converted in a simple sign $(\gamma = +,-)$ in the RHS and the bar represents the opposite element (for instance $\bar e = h$). In Eq. (\ref{eq: coefficients}) we defined the functions
\begin{equation}
    u  = \sqrt{\frac{\Delta}{2\varepsilon}}e^{\frac{1}{2}\mathrm{arccos}\frac{\varepsilon}{\Delta}};\:v  = \sqrt{\frac{\Delta}{2\varepsilon}}e^{-\frac{1}{2}\mathrm{arccos}\frac{\varepsilon}{\Delta}},
\end{equation}
and the phases $\alpha = 2\frac{\varepsilon}{\Delta}\frac{l_S}{\xi_S}$,
$\beta = 2\frac{l_S}{\xi_S}\sqrt{\left(\frac{\varepsilon}{\Delta}\right)^2-1}$,
with $\xi_S =\hbar v_F/\Delta$ the coherence length, $\Delta$ is the superconducting gap and $l_S$ the length of the 2DTI measured from the superconductor as depicted in Fig.  \ref{fig:sketch}.
A similar result for the scattering matrix $S_R$ at the right interface can be obtained. The scattering coefficients can be obtained from Eqs. (\ref{eq: coefficients}) by replacing $\left(r^L_{\alpha,\beta},t^L_{\alpha,\beta}\right)\to \left(r^R_{\beta,\alpha},t^R_{\beta,\alpha}\right)$ and $\phi\to -\phi$.

\subsection{Scattering matrix of the magnetic island}\label{smm}
Following Refs. \onlinecite{nanom, optimal, Bustos, dolcini},
here we compute the scattering matrix $S_M$ describing the edge of the 2DTI in contact with a magnetic island. In order to do this we start by writing the evolution operator 
$\hat{U}^e(x_N,x_0) = \prod_{k=1}^N \hat{U}^e(x_k,x_{k-1}) $ where the superscript $e$ makes reference to the electron part and $N$ indicates the total number of magnetic domains, with
\begin{equation}
\hat{U}^e(x_k,x_{k-1})= \sigma_0\cos\lambda_k+i\vec{n}_k\cdot\hat{\vec{\sigma}}_k\sin \lambda_k,
\end{equation}
where $L_k = x_{k}-x_{k-1}$ is the length of the corresponding magnetic domain. We have introduced  $\lambda_k = L_k\sqrt{\varepsilon^2-\varepsilon_\perp^2}/(\hbar v_F)$, with $\vec{n}_k = \left(i\varepsilon_{k\perp}\sin \theta_k,-i\varepsilon_{k\perp}\cos\theta_k,\varepsilon \right)/\sqrt{\varepsilon^2-\varepsilon_\perp^2}$ and $\theta_k$ is the orientation 
of the domain in the plane of the sample.

The inverse of the evolution operator is the transfer matrix 
\begin{equation}
\left(U^e\right)^{-1} = T^e = \begin{pmatrix}
T^e_{11} & T_{12}^e \\ 
T^e_{21} & T_{22}^e
\end{pmatrix},
\end{equation}
which in turn is related to the scattering matrix as follows
\begin{equation} \label{eq: transfer-scattering magnetic}
S^e_M = \frac{1}{T^e_{22}} \begin{pmatrix}
-T^e_{21} & 1 \\ 
1 & T_{12}^e
\end{pmatrix},
\end{equation}
satisfying the following scattering equation
\begin{equation}
\label{eq: Smelectron}
\left ( c_L^-,c_R^+ \right )^T=S_M^e\left ( c_L^+,c_R^- \right )^T. 
\end{equation} 
A similar relation links the incoming and outgoing holes;
\begin{equation}
\label{eq: Smhole}
\left ( b_L^-,b_R^+ \right )=S_M^h\left ( b_L^+,b_R^- \right ),
\end{equation}
where
$S_M^h(\varepsilon) = -\sigma_z S^{e*}_M(-\varepsilon)\sigma_z$\cite{crepin}.
By combining Eqs. (\ref{eq: Smelectron}) and (\ref{eq: Smhole}), we obtain the scattering matrix for the magnetic island which reads
\begin{equation}
\label{eq: SM}
    S_M = \begin{pmatrix}
r_M & t_M' \\ 
t_M & r_M' 
\end{pmatrix}
=
\begin{pmatrix}
S_{M,11}^e &0  &S_{M,12}^e &0 \\ 
0 & S_{M,11}^h &0 & S_{M,12}^h \\ 
S_{M,21}^e &0  &S_{M,22}^e &0 \\ 
0 & S_{M,21}^h &0 & S_{M,22}^h
\end{pmatrix},
\end{equation}
satisfying the scattering equation
$$\left ( c_L^-,b_L^-,c_R^+,b_R^+ \right )^T = S_M\left ( c_L^+,b_L^+,c_R^-,b_R^- \right )^T.$$
Here we can see that each sub-matrix take a block diagonal form since that in the magnetic domain an electron can not be converted into a hole nor vice-versa in contrast to the case if the SC-2DTI interface which only allows an electron (hole) to be reflected as a hole (electron) or be transmitted as a QP (QH) (see Eq. (\ref{eq: scattering matrix-sc}) ).

\subsection{Combination of the scattering matrices}\label{comb}
By following Ref.\onlinecite{Datta-book} we combine matrices $S_L$ of Eq. (\ref{eq: SL}) and $S_M$ of Eq. (\ref{eq: SM}) and obtain
\begin{equation}
    S_L \circ S_M = \begin{pmatrix}
r & t' \\ 
t &  r'
\end{pmatrix},
\end{equation}
in which
\begin{align}
&r=r_L+t'_Lr_M\left[\mathbb{1}-r'_Lr_M \right ]^{-1}t_L\nonumber\\
&r'=r'_M+t_M\left[\mathbb{1}-r'_Lr_M \right ] ^{-1}r'_Lt'_M\nonumber\\
&t= t_M\left[\mathbb{1}-r'_Lr_M \right ]^{-1}t_L\nonumber\\
&t'=t'_L\left[\mathbb{1}-r_Mr'_L \right ]^{-1}t'_M,
\end{align}
where $\mathbb{1}$ stands for the $2\times 2$ identity matrix.
Finally, by applying the same procedure but adding $S_R$ we obtain the full scattering matrix of the system
\begin{equation}
    S = S_L\circ S_M \circ S_R.
\end{equation}

\bibliography{main.bib}


\end{document}